\documentclass[11pt]{article}
\usepackage{moriond,epsfig,color,cite}
\def\red{\color{red}}
\def\blue{\color{blue}}

\def\magenta{\color{magenta}}

\def\PV{\hbox{\tiny PV}}
\def\PT{\hbox{\tiny PT}}
\def\MSbar {\hbox{$\overline{\hbox{\tiny MS}}\,$}}
\def\eq#1{Eq.~(\ref{#1})}

\begin{document}
\vspace*{4cm}
\title{A NEW APPROACH TO INCLUSIVE DECAY SPECTRA}

\author{Einan Gardi and Jeppe R. Andersen}

\address{Cavendish Laboratory, University of Cambridge\\
Madingley Road, Cambridge, CB3 0HE, UK}

\maketitle \abstracts{The main obstacle in describing inclusive
decay spectra in QCD --- which, in particular, limits the precision
in extrapolating the measured $\bar{B}\longrightarrow X_s \gamma$
rate to the full phase space as well as in extracting $|V_{ub}|$
from inclusive measurements of charmless semileptonic decays
---  is their sensitivity to the non-perturbative momentum distribution
of the heavy quark in the meson. We show that, despite this
sensitivity, resummed perturbation theory has high predictive power.
Conventional Sudakov--resummed perturbation theory describing the
decay of an on-shell heavy quark yields a divergent expansion.
Detailed understanding of this divergence in terms of infrared
renormalons has paved the way for making quantitative predictions.
In particular, the leading renormalon ambiguity cancels out between
the Sudakov factor and the quark pole mass. This cancellation
requires renormalon resummation but involves no non-perturbative
information. Additional effects due to the Fermi motion of the quark
in the meson can be systematically taken into account through {\em
power corrections}, which are {\em only} important near the physical
endpoint. This way the moments of the $\bar{B}\longrightarrow X_s
\gamma$ spectrum with experimentally--accessible cuts --- which had
been so far just parametrized --- were recently {\em computed} by
perturbative means. At Moriond these predictions were confronted
with new data from BaBar. }

\section{Introduction}

Experimental measurements of inclusive decays, such as $\bar{B}
\longrightarrow X_s \gamma$ and $\bar{B} \longrightarrow X_u l
\bar{\nu}$, cannot be performed over the whole phase space.
Therefore, in order to make precision tests of the Standard Model,
we must have a quantitative understanding of the spectra.

Most of the rate of these decays is associated with jet--like
kinematics, where the invariant mass of the hadronic system is
small. It is well known that this important region is difficult to
describe
theoretically~\cite{Neubert:1993um,Bigi:1993ex,Bosch:2004th}: the
observed spectrum is directly sensitive to soft radiation off the
heavy quark which is hard to disentangle from the ``primordial''
momentum distribution of the heavy quark in the {\em meson}.
Technically, this sensitivity leads to the breakdown of the
perturbative expansion in powers of $\alpha_s(m_b)$ as well as that
of the local Operator Product Expansion (OPE) in powers of
$\Lambda/m_b$.

Let us consider the photon--energy spectrum in $\bar{B}
\longrightarrow X_s \gamma$. The non-perturbative nature of this
distribution is obvious from the fact that the physical endpoint is
near $E_{\gamma}=M_B/2$, where $M_B$ is the meson mass, while at any
order in perturbation theory
--- where the initial state is represented by an on-shell heavy
quark --- the spectrum vanishes for $E_{\gamma}>m_b/2$, where $m_b$
is the quark pole mass.
 In reality the quark is off shell and its Fermi
motion translates into smearing~\footnote{This smearing is related
to the structure of the {\em initial} state, and it is independent
of the appearance of resonances in the final state.} of the spectrum
up to $E_{\gamma}\simeq M_B/2$. Given this physical picture the
perturbative approach appears inappropriate. Consequently, the
predominant strategy has been to parametrize the spectrum based on
available data, rather than to compute it. On the other hand, since
total inclusive rates, as well as their first few moments, are
theoretically described by the OPE~\cite{Blok:1993va,Manohar:1993qn}
--- and in particular, if ${\cal O}(\Lambda^2/m_b^2)$ terms are
neglected,  by perturbation theory
--- one expects just small, computable deviations when moderate
kinematic cuts are applied~\cite{Benson:2004sg}. Progressing to more
stringent cuts, resummed perturbation theory is needed. Eventually
it must be complemented by non-perturbative corrections. We will
show that the predictive power of resummed perturbation theory is
significantly higher than it superficially appears to be.

\section{The divergent nature of the perturbative expansion of spectral moments}

A key ingredient in the perturbative approach to inclusive spectra
is the use of moment space. Considering the $\bar{B} \longrightarrow
X_s \gamma$ photon--energy spectrum in terms of $x\equiv
2E_{\gamma}/m_b$, the perturbative coefficients have singular
structure at the perturbative endpoint, $x=1$. Owing to cancellation
of {\em logarithmic} infrared singularities between real--emission
and virtual corrections, the moments
\begin{eqnarray}
\label{moments} \Gamma_N^{\PT}\,\equiv\, \int dE_{\gamma}
\left(\frac{2E_{\gamma}}{m_b}\right)^{N-1} \frac{d\Gamma^{{
\PT}}}{dE_{\gamma}} =\int_0^1
dx\,x^{N-1}\,\frac{d\Gamma^{\PT}(x)}{dx}
\end{eqnarray}
are infrared-- and collinear--safe. $\Gamma_N^{\PT}$ has a well
defined perturbative expansion to all orders. This expansion,
however, does not converge well. It has: (a) {\bf Sudakov
logarithms}, $\ln N$, that are the finite reminder in the
cancellation of logarithmic infrared singularities. These logarithms
appear at any order in the expansion and dominate the coefficients
at large $N$ (b) {\bf infrared renormalons} that reflect {\em
power--like sensitivity to soft momenta} and induce factorial growth
of the coefficients at large orders that renders the perturbative
expansion non-summable.

\begin{figure}[htb]
\begin{center}
\epsfig{file=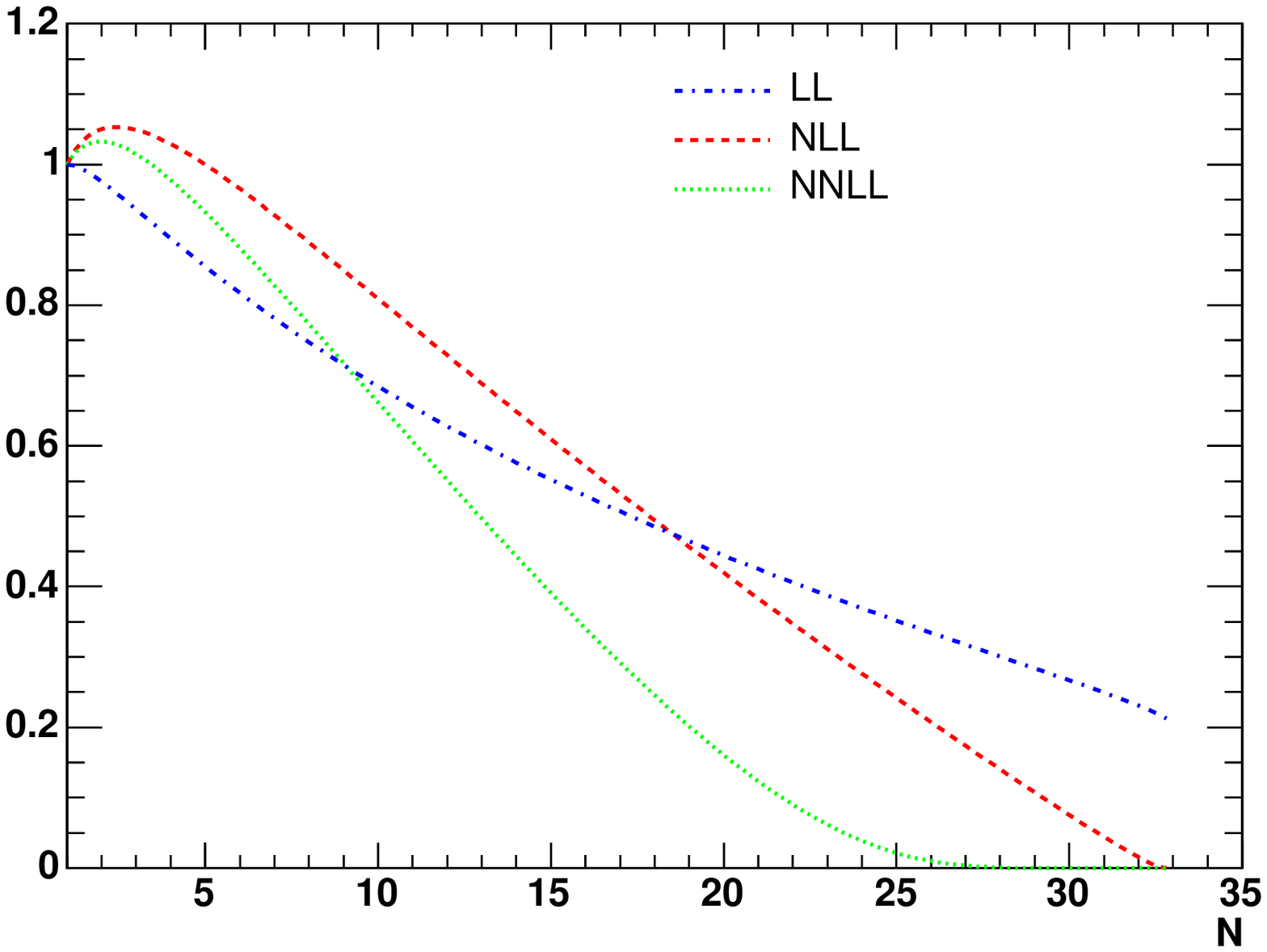,angle=0,width=8.4cm}
\epsfig{file=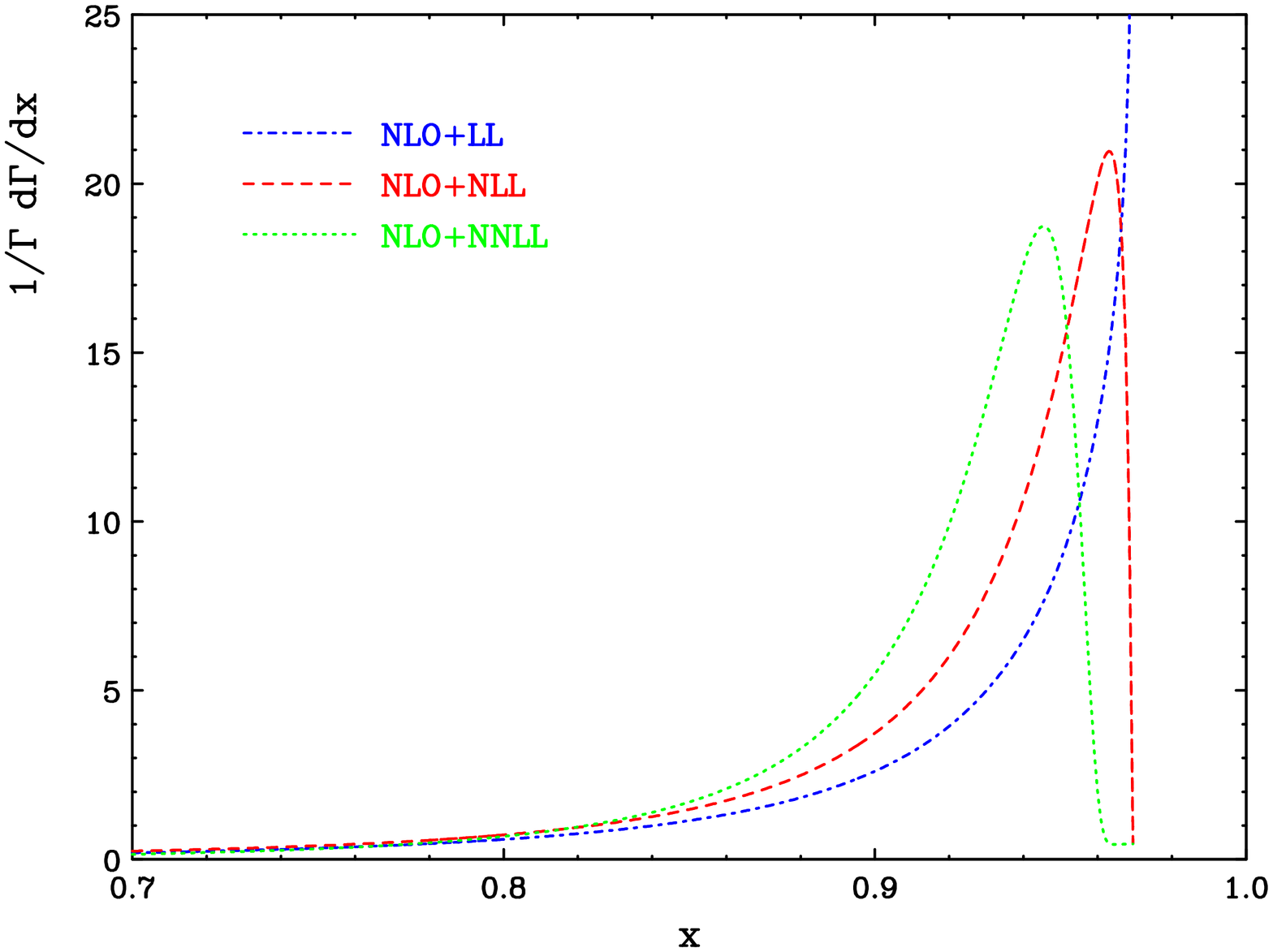,angle=90,width=7.4cm}
\caption{\label{fig:moments_FLA} The moment--space Sudakov factor
${\rm Sud}(N,m_b)$, and the corresponding differential spectrum,
where the exponent is computed to LL, NLL and NNLL accuracy , as
dotdashes, dashes and dots, respectively. In each case the spectrum
is matched to the full NLO result for the magnetic interaction
$O_7$. The curves end at $N\simeq 33$ and at~$x\simeq 0.97$,
respectively, where the Landau singularity in $g_l(\lambda)$
appears.}
\end{center}
\end{figure}

To make quantitative predictions both these sources of large
corrections need to be resummed. The resummation of Sudakov
logarithms is possible thanks to the factorization property of
matrix elements in the soft and the collinear limits. In decay
spectra there are two distinct regions of phase space giving rise to
infrared logarithms~\cite{KS}: {\em soft radiation with momenta of
order $m/N$} from the nearly on-shell heavy quark and {\em
fragmentation of the final--state jet of mass squared $m^2/N$}. In
moment space these contribution factorize:
\begin{equation}
\label{factorization} \Gamma_N^{\PT}= H(m_b) J(m_b^2/N; \mu)
S_{\PT}(m_b/N; \mu)\equiv H(m_b) \times {\rm Sud}(N,m_b),
\end{equation}
up to corrections ${\cal O}(1/N)$. Here $\mu$ is a logarithmic
factorization scale. The Sudakov factor, which sums up the logs to
all orders in $\alpha_s(m_b)$, takes the form of an exponential:
\begin{eqnarray}
\label{Sud} \!\!\!\!{ {\rm Sud}(N,m_b)}\!\!\!\!&=&\!\!\!\!
\exp\left\{-
 \sum_{n=1}^{\infty}\sum_{k=1}^{n+1}{ c_{n,k}}\,{ \ln^{ k} N}
 {\left(\frac{\alpha_s^{\MSbar}(m_b^2)}{\pi}\right)}^{n}
 \right\}\,=\,
\exp\left\{ \sum_{l=0}^{\infty}
g_l(\lambda)\left(\frac{\alpha_s^{\MSbar}(m_b^2)}{\pi}\right)^{l-1}\right\}\!,
\end{eqnarray}
where $\lambda\equiv\left(\beta_0
{\alpha_s^{\MSbar}(m_b^2)}/{\pi}\right)
 \ln N$. Note that ${ {\rm Sud}(N,m_b)}$ depends on the quark
mass only through the argument of the coupling and that it does not
depend on any factorization scale. The first three towers of
logarithms in the exponent, i.e. $c_{n,n+1}$ --- leading logs (LL),
$c_{n,n}$ --- next--to--leading logs (NLL) and $c_{n,n-1}$
--- next--to--next--to--leading logs (NNLL) are
known exactly~\cite{AG,Gardi:2005yi}. The numerical values of the
first few coefficients (for $N_f=4$) are given in Table
\ref{full_coeff}. Obviously, there is problem: the coefficients
increase so fast with increasing powers of $\alpha_s(m_b)$
(corresponding to subleading logarithms) that the sum in the
exponent cannot be determined. The result of conventional Sudakov
resummation with fixed logarithmic accuracy (see Section 2.1 in
Ref.~\cite{AG}) is shown in Fig.~\ref{fig:moments_FLA}. It does not
reach
 perturbative stability.
\begin{table}[h]
 {\small
\begin{eqnarray*}
\begin{array}{rrrrrrrrrr}
      -1.564 &      0.667 &   {\blue 0}\hspace*{22pt}      &   {\blue 0}\hspace*{22pt}      &   {\blue 0}\hspace*{22pt}      &   {\blue 0}\hspace*{22pt}      &   {\blue 0}\hspace*{22pt}      &   {\blue 0}\hspace*{22pt}      \\
     3.837 &      -0.078 &      1.389 &   {\blue 0}\hspace*{22pt}      &   {\blue 0}\hspace*{22pt}      &   {\blue 0}\hspace*{22pt}      &   {\blue 0}\hspace*{22pt}      &   {\blue 0}\hspace*{22pt}      \\
 {\red ?}\hspace*{22pt}        &     20.579 &      6.339 &      3.376 &   {\blue 0}\hspace*{22pt}      &   {\blue 0}\hspace*{22pt}      &   {\blue 0}\hspace*{22pt}      &   {\blue 0}\hspace*{22pt}      \\
 {\red ?}\hspace*{22pt}        & {\red ?}\hspace*{22pt}        &    116.464 &     33.024 &      9.042 &   {\blue 0}\hspace*{22pt}      &   {\blue 0}\hspace*{22pt}      &   {\blue 0}\hspace*{22pt}      \\
 {\red ?}\hspace*{22pt}        & {\red ?}\hspace*{22pt}        & {\red ?}\hspace*{22pt}        &    597.221 &    138.600 &     25.955 &   {\blue 0}\hspace*{22pt}      &   {\blue 0}\hspace*{22pt}      \\
 {\red ?}\hspace*{22pt}        & {\red ?}\hspace*{22pt}        & {\red ?}\hspace*{22pt}        & {\red ?}\hspace*{22pt}        &   2859.284 &    548.170 &     78.492 &   {\blue 0}\hspace*{22pt}      \\
 {\red ?}\hspace*{22pt}        & {\red ?}\hspace*{22pt}        & {\red ?}\hspace*{22pt}        & {\red ?}\hspace*{22pt}        & {\red ?}\hspace*{22pt}        &  13141.289 &   2129.058 &    247.233 \\
 {\red ?}\hspace*{22pt}        & {\red ?}\hspace*{22pt}        & {\red ?}\hspace*{22pt}        & {\red ?}\hspace*{22pt}        & {\red ?}\hspace*{22pt}        & {\red ?}\hspace*{22pt}        &  58941.217 &   8238.359 \\
 {\red ?}\hspace*{22pt}        & {\red ?}\hspace*{22pt}        & {\red ?}\hspace*{22pt}        & {\red ?}\hspace*{22pt}        & {\red ?}\hspace*{22pt}        & {\red ?}\hspace*{22pt}        & {\red ?}\hspace*{22pt}        &  260391.559 \\
\end{array}
\end{eqnarray*}
} \vspace*{-10pt} \caption{\label{full_coeff}The coefficients
$c_{n,k}$ in \eq{Sud} that are known exactly (NNLL accuracy). Rows
(top to bottom) and columns (left to right) correspond to increasing
powers of $n$ (power of $\alpha_s(m)$) and $k$ (power of $\ln N$),
 respectively. The three diagonals are LL, NLL and NNLL, respectively.}
\end{table}
\begin{table}[h]
{\small
\begin{eqnarray*}
\begin{array}{rrrrrrrrrr}
  -1.56 &       0.67 &       {\blue 0 \hspace*{15pt}}  &       {\blue 0 \hspace*{15pt}}  &       {\blue 0 \hspace*{15pt}}  &
    {\blue 0 \hspace*{15pt}}  &       {\blue 0 \hspace*{15pt}}   &{\blue 0 \hspace*{15pt}}  \\
   1.24 &       0.90 &       1.39 &       {\blue 0 \hspace*{15pt}}  &       {\blue 0 \hspace*{15pt}}  &       {\blue 0 \hspace*{15pt}}
   &       {\blue 0 \hspace*{15pt}}  &       {\blue 0 \hspace*{15pt}}  \\
        61.17 &      28.32 &       8.28 &       3.38 &       {\blue 0 \hspace*{15pt}}  &       {\blue 0 \hspace*{15pt}}  &       {\blue 0 \hspace*{15pt}}  &       {\blue 0 \hspace*{15pt}}  &      \\
      1096.06 &     515.20 &     166.25 &      34.89 &       9.04 &       {\blue 0 \hspace*{15pt}}  &       {\blue 0 \hspace*{15pt}}  &      {\blue 0 \hspace*{15pt}} \\
     20399.23 &   10078.43 &    3231.40 &     793.25 &     131.33 &      25.95 &       {\blue 0 \hspace*{15pt}}  &      {\blue 0 \hspace*{15pt}}   \\
    444615.21 &  221481.03 &   73268.94 &   17791.58 &    3514.66 &     482.12 &      78.49 &       {\blue 0 \hspace*{15pt}}  \\
    11342675.74 &  5665794.49 &  1883129.50 &  468180.33 &   91361.30 &   15080.79 &    1768.50 &     247.23 \\
    334032127.30 &  166960507.50 &  55609620.17 &  13867704.58 &  2760946.21 &  449959.01 &   63745.75 &    6532.65 \\
\end{array}
\end{eqnarray*}}
\vspace*{-10pt} \caption{\label{large_beta0_coeff}The coefficients
$c_{n,k}$ in \eq{Sud} computed in the large--$\beta_0$ limit.
Columns and rows are as in Table~\ref{full_coeff}. }
\end{table}

To understand the reason for this divergent behavior it is useful to
consider the large--$\beta_0$ limit, where {\em all} the
coefficients in the exponent can be computed. The numerical values
of the first few coefficients (for $N_f=4$) in this approximation
are given in Table~\ref{large_beta0_coeff}. We observe that: (a) the
result is similar to the full QCD matrix of Table~\ref{full_coeff};
(b) further subleading logarithms have yet larger coefficients. We
therefore deduce that the divergent behavior in
Table~\ref{full_coeff} is related with running--coupling effects.
This is a general
phenomenon~\cite{DGE_thrust,Gardi:2001di,CG,GR,BDK}: the
large--order behavior of Sudakov exponents is dominated by infrared
renormalons. The increasing coefficients of subleading logarithms
build up factorial growth that renders the series non-summable.
Since in the case under consideration the divergent behavior hits
already at low orders it seems that any attempt to use this
expansion is doomed to fail.

\section{Taming the divergence by Dressed Gluon Exponentiation (DGE)}

It was recently shown~\cite{AG} that, despite the divergent behavior
described above, resummed perturbation theory \`a la DGE {\em does}
lead to quantitative predictions for the $\bar{B} \longrightarrow
X_s \gamma$ spectrum:
\begin{eqnarray}
\label{inv_Mellin} \hspace*{-10pt}
\frac{d\Gamma(E_{\gamma})}{dE_\gamma} = \frac{M_B}{2}
\int_{c-i\infty}^{c+i\infty}\frac{dN}{2\pi i} \,\Gamma_N\,
\left(\frac{2E_{\gamma}}{M_B}\right)^{-N} \simeq \frac{m_b^{{\red
\PV}}}{2}\, \int_{c-i\infty}^{c+i\infty}\frac{dN}{2\pi i}\,
\,\Gamma_N^{{ \PT,{\red \PV}}}\, \left(\frac{2E_{\gamma}}{m_b^{{\red
\PV}}}\right)^{-N},
\end{eqnarray}
where the first inverse--Mellin integral is written in terms of
physical spectral moments:
\begin{equation}
\Gamma_N\equiv \int dE_{\gamma}
\left(\frac{2E_{\gamma}}{M_B}\right)^{N-1}
\frac{d\Gamma}{dE_{\gamma}}=\Gamma_N^{{\PT,{\red \PV}}} \,{\rm
e}^{-{(N-1) \bar{\Lambda}_{{\red \PV}}}/{M_B}}\,{{\cal F}((N-1)
\Lambda/M_B)}, \label{Gamma_N_phys}
\end{equation}
with $\bar{\Lambda}_{{\red \PV}}\equiv M_B-m_b^{{\red \PV}}$, and
$\Gamma_N^{{ \PT,{\red \PV}}}$ corresponds to the perturbative
moments of \eq{moments}, where the renormalon ambiguity is
regularized using Principal Value (PV) Borel summation. In the
second inverse--Mellin integral in \eq{inv_Mellin} we assumed ${\cal
F}\simeq 1$, an approximation that is valid away from the endpoint.
Computing the spectrum by resummed perturbation theory involves:
\begin{description}
\item{(a)\,} Defining the Sudakov exponent in $\Gamma_N^{\PT}$
(as an analytic function of $N$ in the complex $N$ plane) using PV
Borel summation:
\begin{eqnarray}
\label{Sud_DGE} &&  \left.{\rm Sud}(N,m_b)\right\vert_{{\red \PV}}
=\exp \Bigg\{ \frac{C_F}{\beta_0}\,{\red {\rm PV}}\int_0^{\infty}
\frac{du}{u} T(u) \left(\frac{\Lambda^2}{m_b^2}\right)^{u} \times
\nonumber \\&& \hspace*{110pt}\bigg[ B_{\cal
S}(u)\Gamma(-2u)(N^{2u}-1)\!-\! B_{\cal
J}(u)\Gamma(-u)(N^{u}-1)\bigg] \Bigg\},
\end{eqnarray}
where $T(u)\equiv (u\delta)^{u\delta }{\rm
e}^{-u\delta}/\Gamma(1+u\delta )$ with
$\delta\equiv\beta_1/\beta_0^2$ and where the Borel transforms of
the quark distribution (soft) and the jet anomalous dimensions,
$B_{\cal S}(u)$ and $B_{\cal J}(u)$, respectively, are defined in
Section~2.2 in Ref.~\cite{AG}.
\item{(b)\,} Applying the {\em same prescription} to the quark pole mass
$m_b$ in \eq{inv_Mellin} --- or, equivalently to $\bar{\Lambda}$ in
\eq{Gamma_N_phys} --- when computing it from the short distance mass
(e.g. from $m_b^{\MSbar}$).
\item{(c)\,} Performing an inverse--Mellin integral according to
\eq{inv_Mellin}.
\end{description}

In general, the calculation of the Sudakov exponent by DGE differ
from conventional Sudakov resummation (truncation of the sum over
$l$ in \eq{Sud}) in two respects. First, on the purely perturbative
level, by using \eq{Sud_DGE} and incorporating the exact analytic
results for the anomalous dimensions in the large--$\beta_0$
limit~\cite{BDK},
\begin{eqnarray}
\label{B_DJ_large_beta0}
B_{\cal S}(u)&=&{\rm e}^{\frac53 u}(1-u)\,+\,{\cal O}(1/\beta_0 ),\nonumber \\
B_{\cal J}(u)&=&\frac{1}{2}\,{\rm e}^{\frac53 u}
\left(\frac{1}{1-u}+\frac{1}{1-u/2}\right) \frac{\sin\pi u}{\pi
u}\,+\,{\cal O}(1/\beta_0 ),
\end{eqnarray}
the resummation of running--coupling contributions is performed to
all orders (rather than to a given logarithmic accuracy) making the
result renormalization--group invariant~\cite{DGE_thrust}. The
power--like infrared sensitivity, which shows up in \eq{Sud} through
the divergence of the series, explicitly appears in \eq{Sud_DGE} as
Borel singularities at integer and half integer values of $u$.  This
brings us to the second difference, which is an important advantage
of DGE, namely having a definite prescription to make power--like
separation between perturbative and non-perturbative contributions.
The PV prescription we choose guarantees that the perturbative
moments are real valued: $\left.{\rm Sud}(N,m_b)\right\vert_{{\red
\PV}}=\left[\left.{\rm Sud}(N^{\magenta *},m_b) \right\vert_{{\red
\PV}}\right]^{\magenta *}$. This choice also defines the
non-perturbative power corrections.

Another key property of this calculation is the exact
cancellation~\cite{BDK} of the leading, $u=1/2$ infrared renormalon
ambiguity~\footnote{This cancellation concerns all the moments and
it comes on top of the cancellation of the $u=1/2$ renormalon in the
total rate~\cite{BBZ}.} between the Sudakov factor and the quark
pole mass in \eq{inv_Mellin}. This means that the infrared
sensitivity, which severely limits the accuracy of the calculation
in a conventional perturbative approach
(Fig.~\ref{fig:moments_FLA}), is nothing else but an artifact of
perturbation theory --- specifically, of using an on-shell heavy
quark in the initial state --- and it is fully removed if renormalon
resummation is applied in both the Sudakov factor and the pole mass.

To understand this cancellation consider the process--independent
definition of the quark distribution in the meson as the Fourier
transform of $\left<{ B(P_B)}\left\vert
\left[\bar{\Psi}(y)\gamma^+\,{ \Phi_y(0,y)}
\,\Psi(0)\right]_{\mu}\right\vert {B(P_B)}\right>$ with respect to
$y^{-}$ where $y$ is a lightlike vector,  $\Phi_y(0,y)$ is a Wilson
line in this direction, and $\mu$ is a (dimensional--regularization)
ultraviolet renormalization scale. Let us compare this distribution
to its perturbative counterpart, namely the quark distribution in an
on-shell heavy quark defined by the Fourier transform of
$\left<{b(p_b)}\left\vert \left[\bar{\Psi}(y)\gamma^+\,{
\Phi_y(0,y)} \,\Psi(0)\right]_{\mu} \right\vert {b(p_b)}\right>$.
While the former is unambiguous, the latter involves an on-shell
heavy quark external state, which is unphysical. While the moments
of the latter distribution are infrared-- and collinear--safe as far
as logarithms are concerned, their infrared sensitivity renders them
ambiguous at power accuracy. The large--$N$ limit corresponds to the
replacements $iP_B^+y^- \longrightarrow N$ and $ip_b^+y^-
\longrightarrow N$ in the two matrix elements, respectively. One
then obtains the following relation between the two soft
functions~\cite{BDK}:
\begin{eqnarray}
\label{Soft_relation} {S({M_B/N};{\mu})}=\, \underbrace{{S_{{\PT}}({
m_b/N};{\mu})}\,\, {\rm e}^{-{(N-1){\bar{\Lambda}}}/{M_B}}}_{{\rm
leading \,\,renormalon \,\,cancels\,\, out}}\, {{\cal F}((N-1)
\Lambda/M_B)},
\end{eqnarray}
where power corrections on the soft scale are split into two
categories: kinematic ones, which involve $\bar{\Lambda}\equiv
M_B-m_b$ and therefore strongly depend on the quark--mass
definition, and dynamical ones (${\cal F}$), which describe the
Fermi motion of the quark in the meson. The OPE
shows~\cite{Neubert:1993um,Bigi:1993ex} that the latter begin at
${\cal O}\Big(((N-1) \Lambda/M_B)^2\Big)$. As indicated in
\eq{Soft_relation} the cancellation of the leading renormalon
ambiguity involves only the kinematic term and therefore it can be
realized without using any non-perturbative input on the quark
distribution in the meson. An explicit calculation of the quark
distribution in an on-shell heavy quark in the large--$\beta_0$
limit~\footnote{Such calculations were performed in a
process--specific context~\cite{BDK} for $\bar{B} \longrightarrow
X_s \gamma$ and $\bar{B} \longrightarrow X_u l \bar{\nu}$ as well as
using the process--independent definition of the quark distribution
function~\cite{Gardi:2005yi}.} shows that the leading renormalon
ambiguity in the exponent in ${S_{{\PT}}({ m_b/N};{\mu})}$ is indeed
equal in magnitude and opposite in sign to the one in the pole
mass~\cite{Beneke:1994sw,Bigi:1994em}, confirming the cancelation in
\eq{Soft_relation}.

Returning to the observable, the replacement of the perturbative
soft function $S_{{\PT}}(m_b/N;\mu)$ in \eq{factorization} by
$S({M_B/N};{\mu})$ of \eq{Soft_relation}, corresponding to the quark
distribution is the {\em meson}, yields the physical spectral
moments defined in \eq{Gamma_N_phys}:
\begin{equation}
\Gamma_N \,=\, H(m_b)\, \underbrace{{\rm Sud}(N,m_b)\,\,{\rm
e}^{-{(N-1){\bar{\Lambda}}}/{M_B}}}_{{\rm leading \,\,renormalon
\,\,cancels\,\, out}}\, {{\cal F}((N-1) \Lambda/M_B)}.
\end{equation}
These moments are free of the leading renormalon ambiguity. Given
our limited knowledge of dynamic power corrections, as a first
approximation it is sensible to neglect them altogether by setting
${\cal F}\simeq 1$. This leads to the second inverse--Mellin formula
in \eq{inv_Mellin}, which is nothing but resummed perturbation
theory.

\begin{figure}[t]
\begin{center}
\epsfig{file=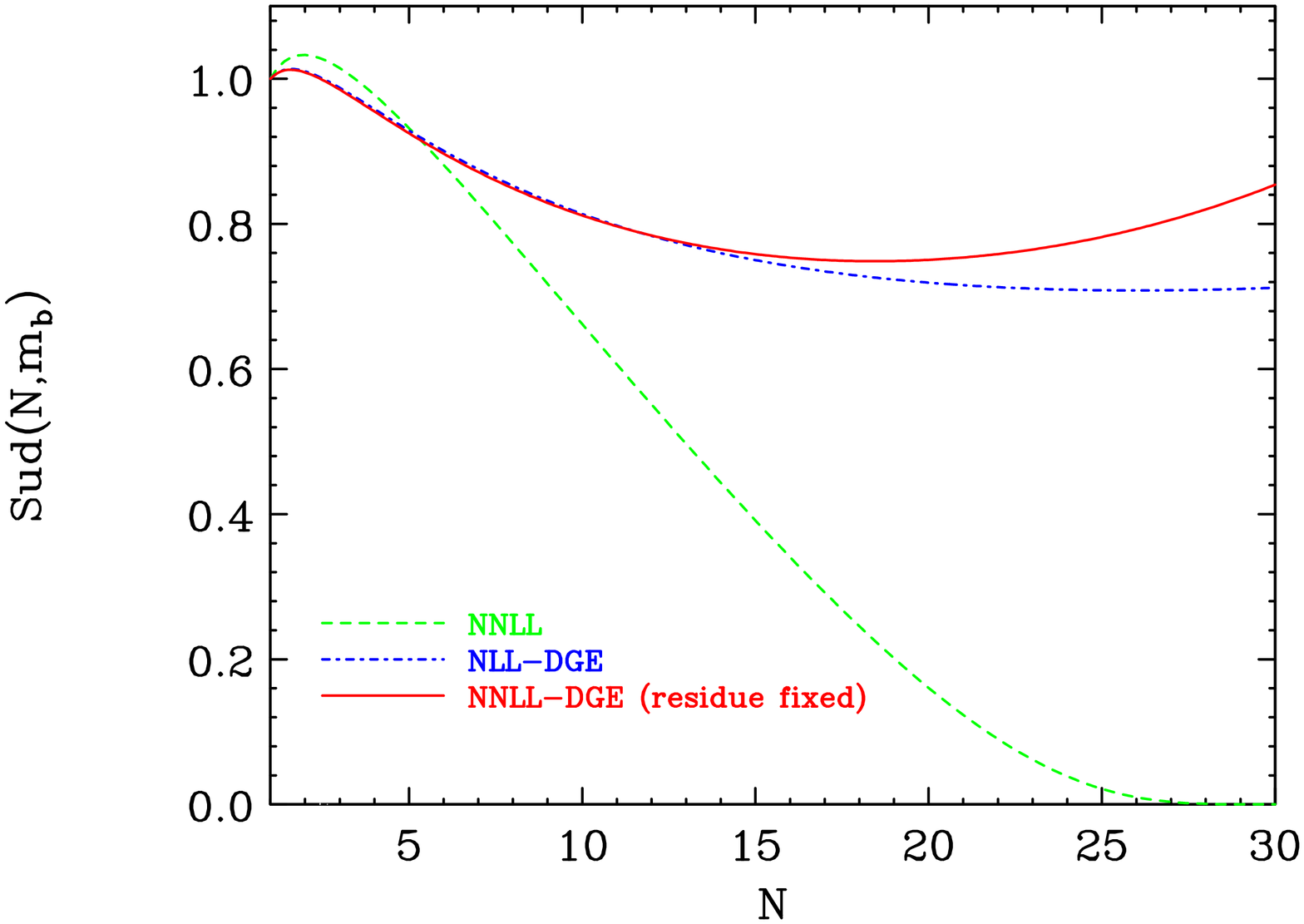,angle=90,width=7.9cm}
\epsfig{file=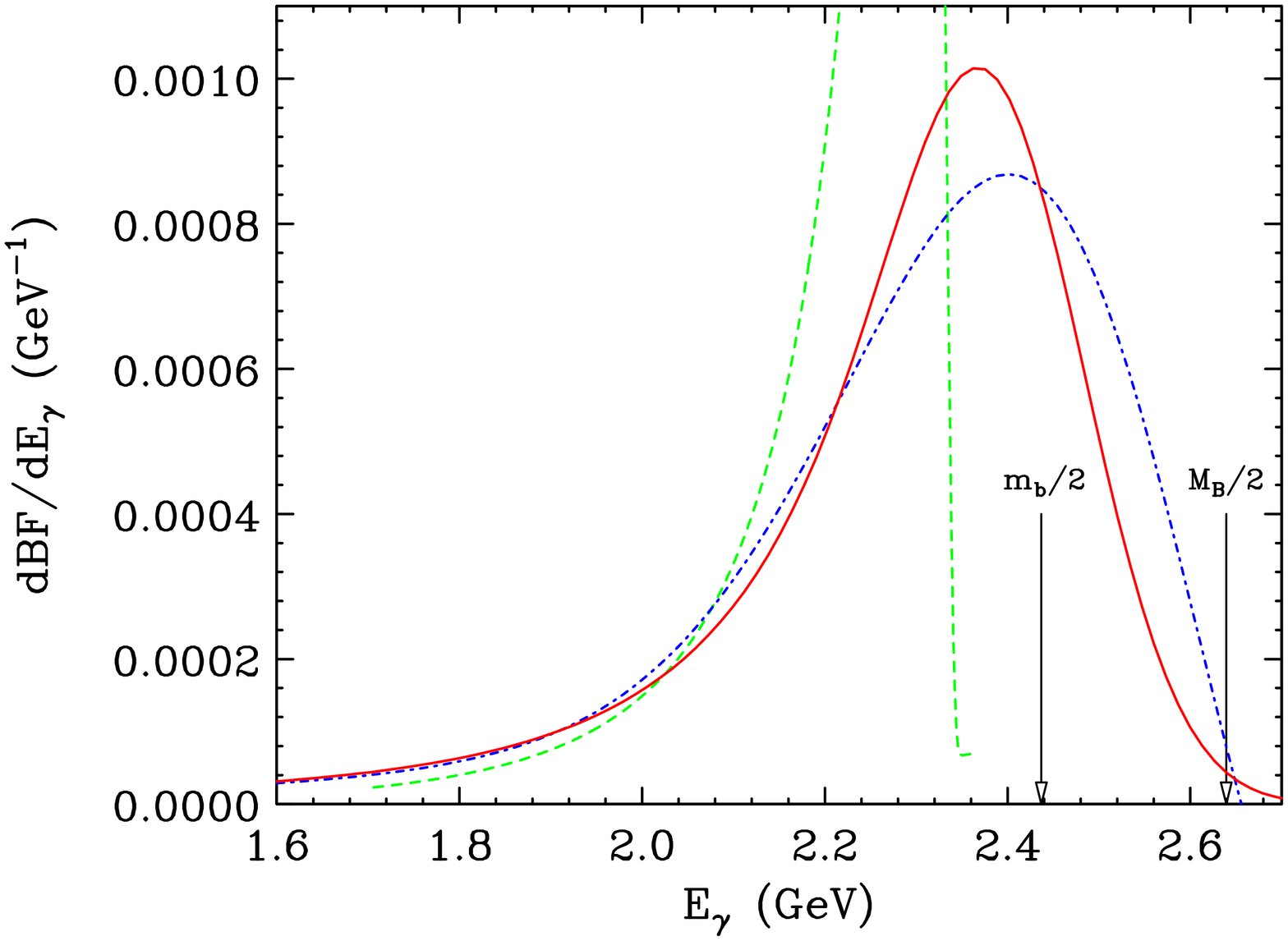,angle=90,width=7.9cm}
\caption{\label{fig:Models} The moment--space Sudakov factor ${\rm
Sud}(N,m_b)$ (left), and the corresponding differential spectrum in
$b\longrightarrow X_s \gamma$ (right) as computed by DGE with
perturbative expansions of NLL (dotdashed blue line) and NNLL (full
red line) accuracy. In the latter also the residue at $u=1/2$ is
fixed. The DGE spectra are both matched with the full NLO. For
comparison the conventional NNLL result is also shown (dashed green
line). }
\end{center}
\end{figure}

In practice, the application of DGE requires evaluating the integral
in~\eq{Sud_DGE} despite incomplete knowledge of the anomalous
dimensions of both the soft and the jet functions. What we currently
know about the exponent includes~\cite{AG}:
\begin{itemize}
\item{} The analytic expressions in the large--$\beta_0$
limit~\cite{BDK} for $B_{\cal S}(u)$ and $B_{\cal J}(u)$ given by
\eq{B_DJ_large_beta0}.
\item{} NNLO (i.e. ${\cal O}(u^2)$) perturbative
expressions in QCD~\cite{AG,BDK} for both $B_{\cal
S}(u)$~\cite{KM,Gardi:2005yi} and $B_{\cal J}(u)$~\cite{GR}.
\item{} The normalization of the leading renormalon
corresponding to the value of $B_{\cal S}(u)$ at $u=1/2$. This value
was determined~\cite{AG}, using the cancellation, based on the
$u=1/2$ renormalon residue in the pole mass, which has been
computed\cite{Pineda:2001zq,AG} with a few percent accuracy from the
NNLO perturbative expansion of the ratio between the pole mass and
$m_b^{\MSbar}$, taking into account the known structure of the
singularity~\cite{Beneke:1994rs}.
\end{itemize}
This information is sufficient~\cite{AG} to constrain the moments up
to $N\sim 20$ fairly well. This is demonstrated in
Fig.~\ref{fig:Models}. The most important source of remaining
uncertainty arises from the unknown behavior of $B_{\cal S}(u)$ away
from the origin, and in particular, for $u>1/2$.
The stability reflected in Fig.~\ref{fig:Models} stands in sharp
contrast with the situation we encountered at fixed logarithmic
accuracy (Fig.~\ref{fig:moments_FLA}), where the leading renormalon
is untamed.

Looking at Fig.~\ref{fig:Models} one immediately observes that {\em
the support properties have changed}. While at any order in
perturbation theory the result is identically zero for
$E_{\gamma}>m_b/2$, upon taking the PV in \eq{Sud_DGE} the
distribution becomes non-zero there. The $E_{\gamma}=m_b/2$ boundary
is a direct consequence of the assumption that the initial $b$ quark
is on-shell. This inherent limitation of perturbation theory is
removed upon defining the sum of the series in moment space by PV.
Remarkably, the PV result tends to zero near the physical endpoint.
This, together with the observed stability, suggests that in this
formulation the dynamical power corrections associated with the
Fermi motion (${\cal F}$), which were neglected here, are indeed
small.

\section{Comparison with data}

\begin{figure}[t]
\begin{center}
\epsfig{file=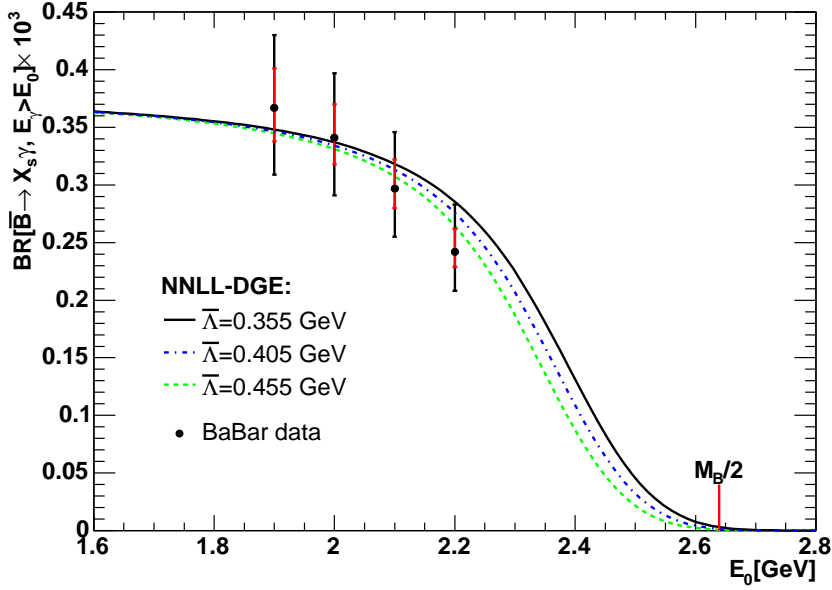,angle=0,width=12cm}
\caption{\label{BF} The partial $\bar{B}\longrightarrow X_s \gamma$
BF  with $E_\gamma>E_0$ as computed by DGE compared with new data
from BaBar. Inner and total error bars on the data points show
systematic and statistical plus systematic errors (added linearly),
respectively. The three theory curves shown (indicated on the plot
by the corresponding values for $\bar{\Lambda}$) correspond to
different input values for $m_b^{\MSbar}$ that roughly cover the
uncertainty range on this parameter. Note that the overall
normalization of the theoretical BF, which is fixed here to its
central value of ${\rm BF} (E_\gamma>E_0=m_b/20)=3.73 \cdot
10^{-4}$, has $\sim 10\%$ uncertainty. }
\end{center}
\end{figure}
Experimentally--favored observables are the partial branching
fraction (BF) for $E_{\gamma}> E_0$,
\begin{equation}
{\rm BF} \equiv  \frac{\Gamma_{\bar{B}\longrightarrow X_s
\gamma}(E_\gamma>{  E_0})}{\Gamma_{\bar{B}\longrightarrow {\rm
anything}}};\qquad\quad \Gamma(E_\gamma>{ E_0})\equiv {\displaystyle
\int_{{ E_0}}dE_{\gamma}\,\frac{d\Gamma(E_{\gamma})}{dE_{\gamma}}},
\end{equation}
where the cut value $E_0$ is above 1.8 GeV, as well as the first few
spectral moments defined over similarly limited range of photon
energies:
\begin{eqnarray}
\label{average_energy} \hspace*{-30pt}{ \Big<}E_{\gamma}{
{{\Big{>_{{\!}_{E_{\gamma}>{ E_0}}}}}}} &\equiv&
\frac{1}{\Gamma(E_\gamma>{ E_0})}\,{\displaystyle \int_{{  E_0}}
dE_{\gamma}\, { \frac{d\Gamma(E_{\gamma})}{dE_{\gamma}}}\,
E_{\gamma}},\\
\label{higher_moments} \hspace*{-30pt} {  \Big<}\Big({
\Big<}E_{\gamma}{ {\Big{>_{\!_{E_{\gamma}>{  E_0}}}}}}-
E_{\gamma}\Big)^n{ \Big{>_{{\!}_{E_{\gamma}>{  E_0}}}}}&\equiv&
\frac{1}{\Gamma(E_\gamma>{  E_0})}{\displaystyle\int_{{  E_0}}
dE_{\gamma}\, { \frac{d\Gamma(E_{\gamma})}{dE_{\gamma}}}\, \Big({
\Big<}E_{\gamma}{ {\Big{>_{{\!}_{E_{\gamma}>{  E_0}}}}}}-
E_{\gamma}\Big)^n}.
\end{eqnarray}
Measurements of the average energy defined in \eq{average_energy},
and the variance, $n=2$ in \eq{higher_moments}, with $E_0=1.815$ GeV
were published~\cite{Koppenburg:2004fz} by Belle in 2004.

\begin{figure}[t]
\begin{center}
\epsfig{angle=90,file=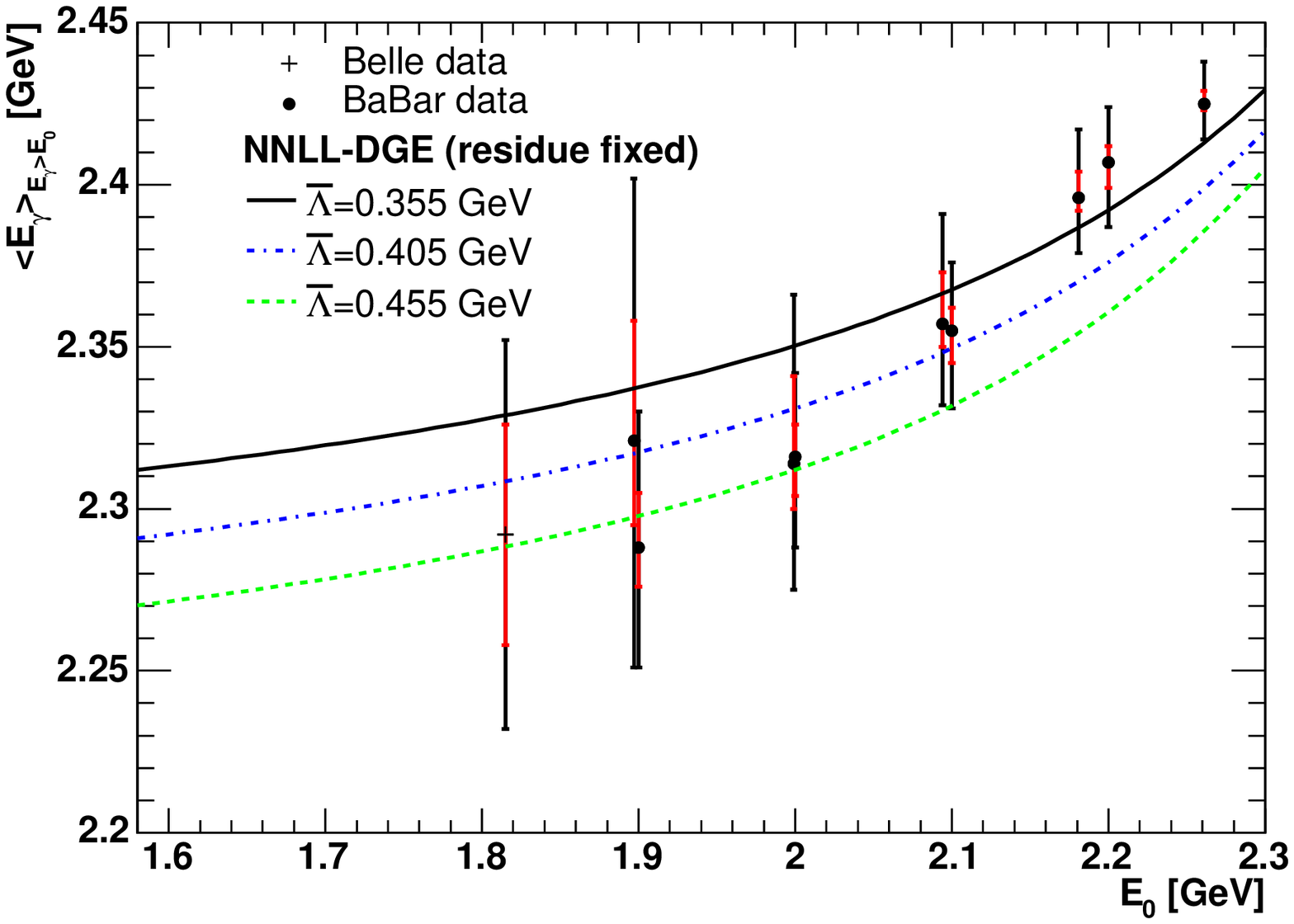,angle=-90,width=7.93cm}
\epsfig{angle=90,file=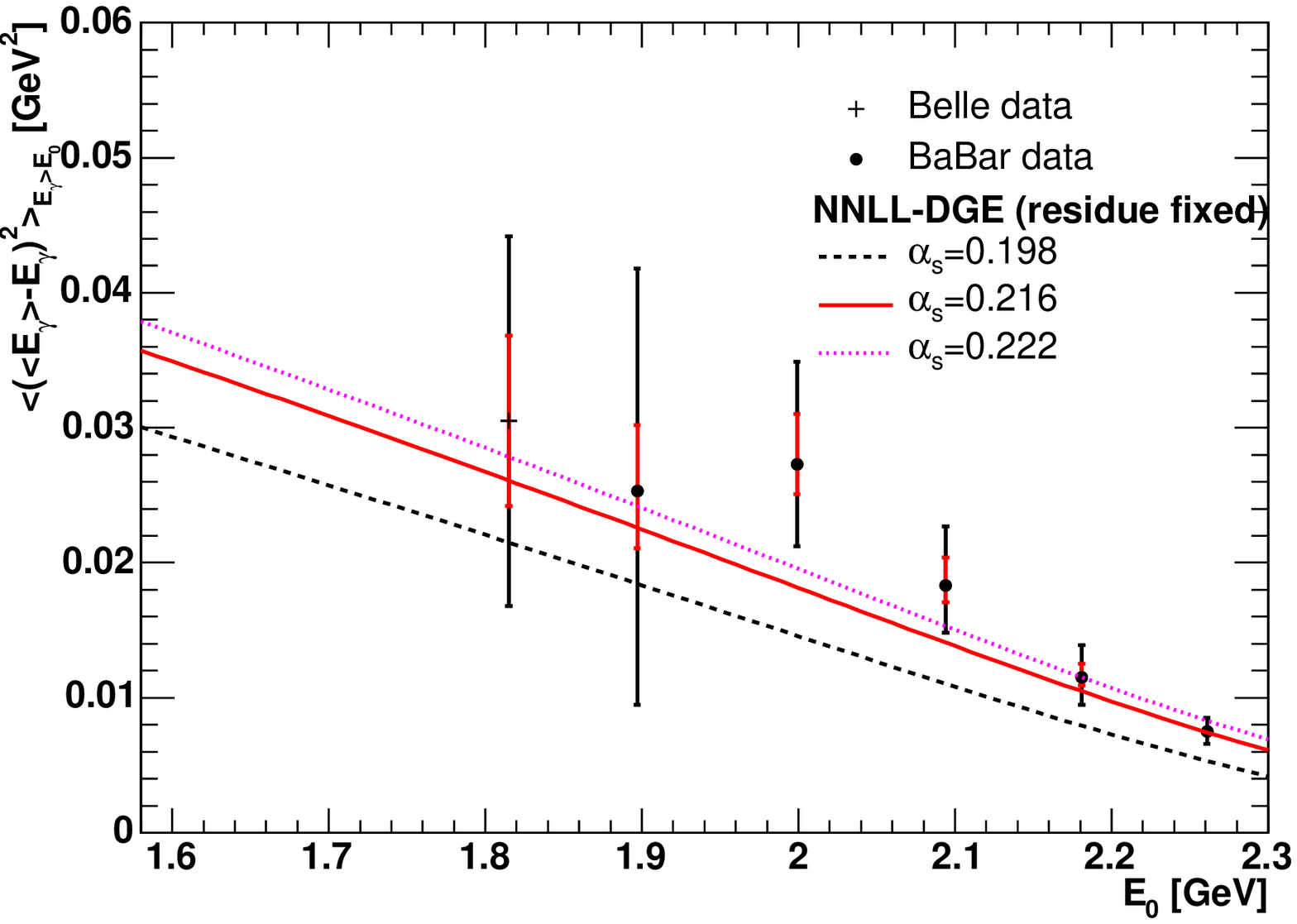,angle=-90,width=7.93cm}
\caption{\label{cut_mom}The first two cut moments of the
$\bar{B}\longrightarrow X_s\gamma$ spectrum:
$\left<E_{\gamma}\right>$ (left) and
$\left<(\left<E_{\gamma}\right>-E_{\gamma})^2\right>$ (right),~as~a
function of the minimum photon energy cut $E_0$, as calculated by
DGE (varying $m_b^{\MSbar}$ and $\alpha_s$ within their error
ranges), compared with data from Belle and BaBar. Inner and total
error bars show systematic and statistical plus systematic errors
(added linearly), respectively. }
\end{center}
\end{figure}

The method presented in the previous section facilitates theoretical
calculation of the $\bar{B}\longrightarrow X_s \gamma$ spectrum, and
thus also evaluation of the cut moments defined above, by purely
perturbative means, so long as power corrections (${\cal F}$) can be
neglected. Since these power corrections are controlled by the soft
scale $m_b/N$, they become increasingly important with increasing
$E_0$. In order to test the theory and, eventually, determine these
power corrections it is therefore useful~\cite{AG} to compare the
theoretical calculation with data as a function of $E_0$. Such
comparison was done for the first time {\em during} the Moriond
meeting, following the talk by J.J.~Walsh~\cite{Walsh} who presented
new (and preliminary) results from Babar; see Figs.~\ref{BF}
and~\ref{cut_mom}.

\section{Conclusions}

We presented a new approach to inclusive decay spectra in QCD and
its application to the $\bar{B}\longrightarrow X_s\gamma$ spectrum.
Our method, Dressed Gluon Exponentiation, incorporates renormalon
resummation as well as Sudakov resummation and uses the PV
prescription to separate between computable perturbative
contributions and non-perturbative power corrections, which become
important near the endpoint.

We have shown that, despite the apparent infrared sensitivity which
shows up as divergence of the perturbative expansion in \eq{Sud},
the predictive power of resummed perturbation theory is high. This
is reflected, for example, in the stability of the DGE result shown
in Fig.~\ref{fig:Models}. A~key property underlying this predictive
power is the exact cancellation of the leading infrared renormalon
ambiguity~\cite{BDK}. This cancellation can be understood as a
manifestation of the infrared-- and collinear--safety of the
physical spectral moments at power accuracy.

In the calculation of the spectrum by \eq{inv_Mellin} the leading
renormalon cancellation is realized by using the same prescription
when defining the sum in the Sudakov exponent and when computing the
quark pole mass which sets the overall energy scale. Renormalon
resummation in both these quantities is of course imperative for
this cancellation to take place. The final result for the spectrum
is prescription independent as far as this leading renormalon is
concerned. Having neglected higher power corrections on the soft
scale, our result does depend on using the PV prescription for
higher renormalon singularities. While we do not assign to the
chosen regularization any physical meaning, we find the PV
prescription particularly useful. Specifically, the PV--regularized
spectrum shares the following key properties with the physical
spectrum: (a) the moments are real valued, (b) the
resummed spectrum smoothly extends {\em beyond the perturbative
endpoint} and, quite remarkably, tends to zero for
$E_{\gamma}=\left(m_b+{\cal O}(\Lambda)\right)/2$, close to $M_B/2$.
This suggest that in this formulation power corrections are indeed
small.

Finally, as shown in Figs.~\ref{BF} and~\ref{cut_mom}, there is good
agreement between the theoretical predictions and data from the B
factories for both the partial BF and the first two spectral moments
over the entire range of cuts where data is available. Fits can
readily be performed to measure $m_b^{\MSbar}$ and to put bounds on
the dynamical power corrections, which were so far neglected in this
approach. Finally, this successful comparison with data shows that
prospects are good for precision determination of $V_{ub}$ from
charmless semileptonic decays using DGE.

\section*{Acknowledgments}

EG wishes to thank Vladimir Braun, Gregory Korchemsky and Arkady
Vainshtein for very interesting discussions. JRA acknowledges the
support of PPARC (postdoctoral fellowship PPA/P/S/2003/00281). The
work of EG is supported by a Marie Curie individual fellowship,
contract number HPMF-CT-2002-02112.


\section*{References}
\begin{thebibliography}{99}

\bibitem{Neubert:1993um}
   M.~Neubert,
   {\em Phys. Rev.} {\bf D49} (1994) 4623; [hep-ph/9312311].
   {\em Phys. Rev.} {\bf D49} (1994) 3392
   [hep-ph/9311325].

\bibitem{Bigi:1993ex}
  I.~I.~Y.~Bigi, M.~A.~Shifman, N.~G.~Uraltsev and A.~I.~Vainshtein,
  {\em Int. J. Mod. Phys.} {\bf A9} (1994) 2467
  [hep-ph/9312359].

\bibitem{Bosch:2004th}
  S.~W.~Bosch, B.~O.~Lange, M.~Neubert and G.~Paz,
  {\em Nucl. Phys.} {\bf B699} (2004) 335
  [hep-ph/0402094].

\bibitem{Blok:1993va}
  B.~Blok, L.~Koyrakh, M.~A.~Shifman and A.~I.~Vainshtein,
  {\em Phys. Rev.} {\bf D49} (1994) 3356 [Erratum-ibid.\ {\bf  D50}
  (1994) 3572] [hep-ph/9307247].

\bibitem{Manohar:1993qn}
  A.~V.~Manohar and M.~B.~Wise,
  {\em Phys. Rev.} {\bf D49} (1994) 1310 [hep-ph/9308246].

\bibitem{Benson:2004sg}
  D.~Benson, I.~I.~Bigi and N.~Uraltsev,
  {\em Nucl. Phys.} {\bf B710}, 371 (2005)
  [hep-ph/0410080].

\bibitem{KS}
  G.~P.~Korchemsky and G.~Sterman,
  {\em Phys. Lett.} {\bf B340}, 96 (1994) [hep-ph/9407344].

\bibitem{AG}
  J.~R.~Andersen and E.~Gardi,
  ``Taming the B $\to$ X/s gamma spectrum by dressed gluon exponentiation,''
  [hep-ph/0502159].

\bibitem{Gardi:2005yi}
  E.~Gardi,
  JHEP {\bf 0502} (2005) 053
  [hep-ph/0501257].

\bibitem{DGE_thrust}
  E.~Gardi and J.~Rathsman,
  {\em Nucl. Phys.}  {\bf B609} (2001) 123 [hep-ph/0103217];
  %
  {\em Nucl. Phys.} {\bf B638} (2002) 243 [hep-ph/0201019].


\bibitem{Gardi:2001di}
  E.~Gardi,
  {\em Nucl. Phys.} {\bf B622} (2002) 365 [hep-ph/0108222].

\bibitem{CG}
  M.~Cacciari and E.~Gardi,
  {\em Nucl. Phys.} {\bf B664} (2003) 299 [hep-ph/0301047].

\bibitem{GR}
  E.~Gardi and R.~G.~Roberts,
  {\em Nucl. Phys.} {\bf B653}, 227 (2003) [hep-ph/0210429].

\bibitem{BDK}
  E.~Gardi,
  JHEP {\bf 0404}, 049 (2004)
  [hep-ph/0403249].

\bibitem{Beneke:1994sw}
  M.~Beneke and V.~M.~Braun,
  {\em Nucl. Phys.}  {\bf B426}, 301 (1994) [hep-ph/9402364].

\bibitem{Bigi:1994em}
  I.~I.~Y.~Bigi, M.~A.~Shifman, N.~G.~Uraltsev and A.~I.~Vainshtein,
  {\em Phys. Rev.} {\bf D50}, 2234 (1994) [hep-ph/9402360].

\bibitem{KM}
  G.~P.~Korchemsky and G.~Marchesini,
  {\em Nucl. Phys.} {\bf B406} (1993) 225
  [hep-ph/9210281].

\bibitem{BBZ}
  M.~Beneke, V.~M.~Braun and V.~I.~Zakharov,
  {\em Phys. Rev. Lett.}  {\bf 73} (1994) 3058
  [hep-ph/9405304].

\bibitem{Pineda:2001zq}
  A.~Pineda,
  JHEP {\bf 0106} (2001) 022 [hep-ph/0105008];
%
  T.~Lee,
  JHEP {\bf 0310}, 044 (2003) [hep-ph/0304185];
%
  G.~Cvetic,
  {\em J.\ Phys.\ G} {\bf 30} (2004) 863 [hep-ph/0309262].

\bibitem{Beneke:1994rs}
  M.~Beneke,
  {\em Phys. Lett.} {\bf B344} (1995) 341 [hep-ph/9408380].

\bibitem{Koppenburg:2004fz}
  P.~Koppenburg {\it et al.}  [Belle Coll.],
  {\em Phys.\ Rev.\ Lett.}  {\bf 93} (2004) 061803 [hep-ex/0403004].

\bibitem{Walsh}
  J.J.~Walsh [BaBar Coll.] ``Semileptonic and EW Penguin Decay
  Results from BaBar'', Talk at the XXXXth Rencontres de Moriond QCD
  and Hadronic Interactions La Thuile, Italy March 12 -- March 19,
  2005.\\
   {\tt http://moriond.in2p3.fr/QCD/2005/Index.html}



\end{thebibliography}
\end{document}